\documentstyle[epsfig,color,12pt]{article}

\include{epsf}

\def\Lem{Lema\^\i tre}

\def\beq{\begin{equation}}
\def\eeq{\end{equation}}

\begin{document}

\centerline{\large\bf Triple-horizon spherically symmetric
spacetime and  holographic principle}

\vskip 0.2in

\centerline{\large\it{Irina~Dymnikova}}

\vskip 0.2in

\centerline{\it Department of Mathematics and Computer Science,
University of Warmia and Mazury,}

\centerline{\it S{\l}oneczna 54, 10-710  Olsztyn, Poland; e-mail:
irina@matman.uwm.edu.pl}

\centerline{\it A.F. Ioffe Physico-Technical Institute,
Politekhnicheskaja 26, St.Petersburg, 194021 Russia}

\vskip0.4in

Honorable Mentioned Essay - Gravity Research Foundation 2012

Submitted to Int. J. Mod. Phys. D

\vskip0.3in

{\bf Abstract}

\vskip0.1in

We present  a family of spherically symmetric spacetimes,
specified by the density profile of a vacuum dark energy, which
have the same global structure as the de Sitter spacetime but the
reduced symmetry which leads to a time-evolving and spatially
inhomogeneous cosmological term.  It connects smoothly two de
Sitter vacua with different values of cosmological constant and
corresponds to anisotropic vacuum dark fluid defined by symmetry
of its stress-energy tensor which is invariant under the radial
boosts.  This family contains a special class distinguished by
dynamics of evaporation of a cosmological horizon which evolves to
the triple horizon with the finite entropy, zero temperature, zero
curvature, infinite positive specific heat, and infinite
scrambling time. Non-zero value of the cosmological constant in
the triple-horizon spacetime is tightly fixed by quantum dynamics
of evaporation of the cosmological horizon.

\newpage

The holographic principle was first formulated \cite{hooft} as the
requirement that the number of independent quantum degrees of
freedom $N_d$ contained in a given spatial volume $V$ is bounded
from above by the surface area of the region, $N_d(V)\leq A/4$. It
led to a conjecture that a physical system can be completely
specified by data stored on its boundary, which obtained the name
of holographic principle \cite{susskind,bousso}.
 This view was strongly supported by the
AdS/CFT duality  between quantum gravity in AdS space and a
certain conformal field theory on its boundary \cite{ads/cft}
which defines quantum gravity non-perturbatively and involves only
physical degrees of freedom admitted by the holographic principle
\cite{suswit}.

\vskip0.1in

Astronomical observations testify for an early inflationary
stage in our universe evolution and provide convincing evidence that now it
is dominated at above 70 \% of its density by a dark energy responsible for
its accelerated expansion due to negative pressure, $p=w\rho$, $w<-1/3$ \cite{DE},
with the best fit $w=-1$ \cite{lambda} corresponding to a positive cosmological
constant $\lambda$, so that we are living in the universe asymptotically de Sitter
in both past and future.

\vskip0.1in

The question of existence of a holographic duality between quantum
gravity in de Sitter space and an Euclidean conformal field theory
on its boundary, DS/CFT correspondence \cite{strominger}, is one
of the most intriguing questions discussed in the literature.
 A holographic description of de Sitter
space requires the symmetries of de Sitter space to act on its
asymptotic boundary as conformal mappings \cite{dyson-susskind}.
It encounters the de Sitter complementarity principle
\cite{dyson-susskind} referred to also as observer complementarity
\cite{verlinde} since it represents the symmetry relating
observations of observers passing out of each other's static
patches \cite{susskind2011}. The key point is that the de Sitter
horizon is observer-dependent \cite{GH}. Each observer sees the
surrounding spacetime as a finite closed cavity bounded by his
horizon with the finite temperature and entropy. The de Sitter
complementarity precludes the existence of the appropriate limits
for the boundary correlators in the DS/CFT correspondence if the
entropy is finite \cite{dyson-susskind}.

An additional trouble  comes from the fact that de Sitter horizon
is a fast scrambler \cite{susskind2011}. The scrambling time for a
quantum system at temperature $T$ is the time needed to thermalize
information, given that the interactions are between bounded
clusters of degrees of freedom \cite{sekino}. A universal bound on
scrambling time $t_*$ for a given entropy $S$ can be estimated as
\cite{susskind2011}
   \beq
   t_* \geq A ~ T^{-1}~ \hbar~  S^{1/2}
                                           \label{scramble}
  \eeq
  where $A$ is a certain constant.
For de Sitter horizon $t_*=T^{-1}l\ln(l/l_P)$ \cite{susskind2011}
where $T$ is the temperature of the horizon $l$. This qualifies it
as a fast scrambler and leads to the conclusion that it cannot be
described by any kind of two-dimensional field theory
\cite{susskind2011}, although there could exist a dual description
based on the matrix quantum mechanics \cite{banks}. However, the
question of consistence with observer complementarity remains open
- different static patches are related by the non-compact ${\cal
O}(4,1)$ symmetry of de Sitter space, and the problem is how to
implement such a symmetry on the matrix degrees of freedom which
describe only a single static patch \cite{susskind2011}. In
\cite{kleban-susskind} the no-go theorem was proved that the
non-compact ${\cal O}(4,1)$ symmetry of de Sitter space  cannot be
realized if the entropy is finite \cite{susskind2011} with the
conclusion that {\it if the entropy is finite then the symmetry of
different causal patches must be broken} \cite{kleban-susskind}.

\vskip0.2in

In this note we present  a family of spherically symmetric
spacetimes, specified by the density profile of a vacuum dark
energy, which has the same global structure as the de Sitter
spacetime but the reduced symmetry which leads to a time-evolving
and spatially inhomogeneous cosmological term. This family
contains a special class distinguished by dynamics of evaporation
of a horizon which evolves to thermodynamically stable state with
an infinite scrambling time.

\newpage

The Einstein cosmological term $\Lambda g_{ik}$ corresponds to
the maximally symmetric stress-energy tensor
   \beq
T^i_k=\rho_{vac}\delta^i_k; ~ ~ ~\rho_{vac}=(8\pi G)^{-1}\Lambda
                                                                \label{maxvac}
   \eeq
 which is responsible for the high symmetry of  de Sitter
 spacetime.
The tensor(\ref{maxvac}) was interpreted  as a Lorentz-invariant
vacuum fluid by \Lem\ as early as in 1934 \cite{lem}, and in 1965
Gliner supplemented this interpretation with the elegant argument:
the medium specified by (\ref{maxvac}) has an infinite set of
  co-moving reference frames, so that an observer cannot in principle measure
  his/her velocity with respect to it \cite{Gliner}, which is the intrinsic property of a
  vacuum \cite{landau}.

The model-independent way to make $\Lambda$ variable without
introducing additional entities is to reduce symmetry of
(\ref{maxvac}) (which enforces $\Lambda =$ const by $T^i_{k;i}=0$)
while keeping its vacuum identity, i.e. Lorentz-invariance in one
or two spatial direction and thus impossibility to fix some
of velocity components. A stress-energy tensor with a reduced
symmetry such that only one or two of
  its spatial eigenvalues coincides with the temporal eigenvalue, represents
  a vacuum dark fluid with the equation of state $p_{\alpha}=-\rho$ in the distinguished
  direction(s) \cite{dg2007}, which makes it intrinsically anisotropic. This immediately promotes
  the cosmological constant $\Lambda$  to a dynamical component $\Lambda^0_0$ of a variable
  (by $T^i_{k;i}=0$) cosmological term
  $\Lambda^i_k=8\pi GT^i_k$  \cite{me2000}.

The spherically symmetric vacuum is specified by \cite{me92}
\beq
T^0_0=T^1_1~  ~ ~ ~ (p_r=-\rho)
                                                 \label{myvac}
\eeq
Transversal pressure is given by
    \beq
    p_{\perp}=-\rho -\frac{r}{2}\rho'
                                                 \label{trans}
    \eeq
 and the metric has the form
\beq
 ds^2=g(r)dt^2-\frac{dr^2}{g(r)}-r^2d\Omega^2
                                                 \label{metric}
\eeq
 with the metric function
\beq
 g(r)=1-\frac{2G{\cal M}(r)}{r};~ ~  ~{\cal
M}(r)=4\pi\int_0^r{\rho(x)x^2dx}
                                                   \label{gfunc}
   \eeq
 For the source terms satisfying the
weak energy condition, regular solutions have obligatory de Sitter
center and monotonically decreasing density $\rho(r)$
\cite{me2002}. Therefore  we can separate in (\ref{gfunc})
  the minimal vacuum density $\rho_{\lambda}$ as the background density.
   If we introduce $\rho(r)=\rho_{d}(r)+\rho_{\lambda}$
  where $\rho_d(r)$ is a dynamical density
  decreasing  from the value at the center $\rho_0=(8\pi G)^{-1}\Lambda$ to zero at infinity,
"cosmological constant" $\Lambda^0_0$
evolves monotonically from $\Lambda+\lambda$  to $\lambda$
\cite{me2002}.
  In the case of two vacuum scales, $\Lambda=8\pi G\rho_{0}$ at the
center, and $\lambda=8\pi G\rho_{\lambda}$  at infinity, geometries are defined by three
parameters: two length scales related to limiting vacuum
densities, $r_0=\sqrt{3/\Lambda}$ and $l=\sqrt{3/\lambda}$, and
the mass parameter $M=4\pi\int_0^{\infty}{\rho_d(r)r^2dr}$.

The curvature scalar
  \beq
{\cal R}=8\pi G(4\rho + r\rho^{\prime})=-32\pi Gp_{\perp}
                                                         \label{curvature}
   \eeq
can have two zero points since $p_{\perp}$ is negative for
$r\rightarrow 0$ and for $r\rightarrow\infty$.

The  behavior of the metric
  function $g(r)$ is dictated by the number of vacuum scales through behavior
   of the transversal pressure $p_{\perp}$ in a source term.
By the Einstein equation \cite{landau}
   \beq
8\pi G p_{\perp} = g''/2 + {g'}/{r}
                                                   \label{trans1}
   \eeq
it determines the maximal number and character of $g(r)$ extrema
which in turn determines the maximal number of spacetime horizons
\cite{bdg}. In the case of two vacuum scales, $p_{\perp}=-\rho$ as
$r\rightarrow 0$ and as $r\rightarrow\infty$, hence $g(r)$ can
have two maxima (one obligatory in de Sitter center $r=0$), one
minimum and not more than three zeros, and spacetime can thus have
not more than three horizons \cite{bdd}. In the 3-horizon case the
metric (\ref{metric}) describes a regular black hole with de
Sitter center in the asymptotically de Sitter space \cite{us97}
shown in Fig.1.
%
\begin{figure}[htp]
\centering \epsfig{file=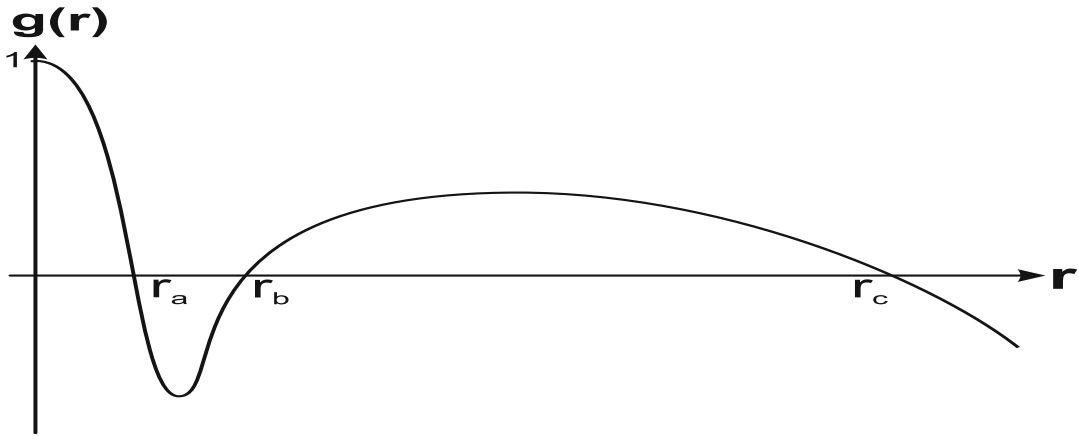,width=14.3cm,height=5.7cm}
\caption{Typical behavior of a metric function $g(r)$ for the case
of three horizons: a black hole horizon $r_b$,  a cosmological
horizon $r_c$, and an internal horizon $r_a$ related to the de
Sitter center. } \label{fig.1}
\end{figure}

For a black hole the mass parameter is confined within  $M_{cr1} <
M < M_{cr2}$. Values $M_{cr1}$ and $M_{cr2}$ depend on the
parameter $q=\sqrt{\Lambda/\lambda}$ \cite{us97}. Double-horizon
configurations $M_{cr1}$ and $M_{cr2}$ represent, respectively, an
extreme black hole and a regular modification of the Nariai
solution. Configurations $M <M_{cr1}$ and $M>M_{cr2}$ correspond
to one-horizon spacetimes (see Fig.2).
\begin{figure}[htp]
\centering
\epsfig{file=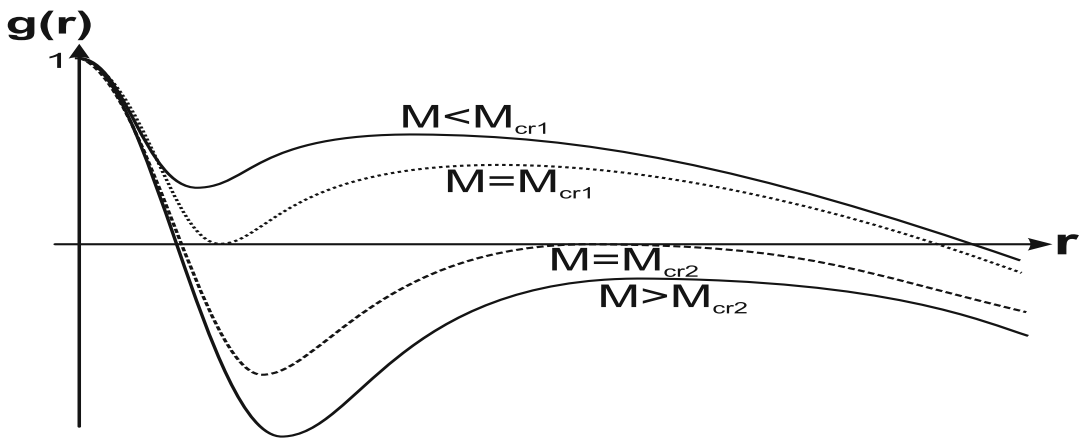,width=14.3cm,height=5.7cm}
    \caption{Metric function for spacetime with de Sitter center
    asymptotically de Sitter as $r\rightarrow\infty$. }
     \label{fig.2}
\end{figure}

\vskip0.1in

Cosmological evolution of an anisotropic fluid is described by
the \Lem\ class models. The static spherically symmetric metric
(\ref{metric}) can always be transformed to the \Lem\ form
\cite{bdd,observers,bdg} with the line element \cite{landau}
  \beq
    ds^2  = d\tau^2 -e^{\nu (R,\tau)} dR^2
                    - r^2 (R,\tau) d\Omega^2,        \label{ds2Lem}
    \eeq
 where $e^{\nu(R, \tau)} = {r'^2}=(\partial r/\partial R)^2$
 for the case when each 3-hypersurface $\tau = const$ is flat  \cite{landau}, which guarantees
 fulfilment of the spatial flatness condition $\Omega=1$ required by  observations.

The function $r(R, \tau)$ is found by integration of the Einstein
equations which gives \cite{ddfg,bdd}
   \beq
    \tau-\tau_0(R)=\int \frac{dr}{\sqrt{2G{\cal M}(r)/r}};~ ~ ~ {\cal
M}(r)=4\pi\int_0^r{\rho(x)x^2dx}   \label{tau-lem}
  \eeq
  and the function $e^{\nu(R, \tau)}$ is found from (\ref{tau-lem}) as
    $e^{\nu(R,\tau)} = 2G {\cal M}(r(R, \tau))/r(R, \tau)
    (\tau_0'(R))^2$ \cite{bdd,observers}.

 For the case of the source terms specified by (\ref{myvac}),
regular cosmological solutions are asymptotically de Sitter at the
early and late time, and can describe evolution from an
inflationary beginning to the late accelerated expansion, guided
by dynamical vacuum dark energy closely related to spacetime
symmetry \cite{bdd,bdg}.

\vskip0.1in

Each one-horizon spacetime has the global structure of de Sitter
spacetime \cite{observers}, but spacetime symmetry is essentially
different. The horizon $r_h$ is not observer-dependent as in the
de Sitter case, but real horizon due to reduced symmetry
(\ref{myvac}) which breaks the ${\cal O}(4,1)$ symmetries by
involving the distinguished center $r=0$.

 An observer in the R-region $0\leq r < r_h$ (a static
causal patch) can detect the Hawking radiation from the
cosmological horizon $r_h$. Dynamics of quantum evaporation of
horizon can tell us which of one-horizon spacetimes
 is the most appropriate to represent evolution from the de Sitter beginning to
the de Sitter end.

 The Gibbons-Hawking temperature on a horizon $r_h$ with
a surface gravity $\kappa_h$ is given by \cite{GH}
   \beq
kT_H=\frac{\hbar}{4\pi c}|g^{\prime}(r_h)| \label{temp}
  \eeq

A general approach for defining thermodynamical variables
in a multi-horizon spacetime with non-zero pressure
was developed by Padmanabhan \cite{padmanabhan}. He considered a
canonical ensemble of spacetime metrics from the class
(\ref{metric}) at the constant temperature of the horizon
determined by the periodicity of the Euclidean time in the
Euclidean continuation of the Einstein action. The partition
function calculated as the path integral sum \cite{padmanabhan},
can be written as
   \beq
Z(kT_h)=Z_0\exp\left[\frac{1}{4}\left(4\pi r_h^2\right)
-\frac{1}{kT_h}\left(\frac{|g'|}{g'}\frac{r_h}{2}\right)\right]
\propto \exp\left[S(r_h)-\frac{E(r_h)}{kT_h}\right]
                                                                      \label{Z}
   \eeq
which gives \cite{padmanabhan} (in the geometrical units
$c=G=\hbar=1$)
  \beq
S=\pi r_h^2;  ~~ E=\frac{|g'|}{g}
\left(\frac{A_h}{16\pi}\right)^{1/2}=\frac{|g'|}{g}\frac{r_h}{2}
                                                                    \label{energy}
   \eeq
where $S$ is the  entropy, $E$ is the thermodynamical energy, and
$A_h$ is the horizon area.

A specific heat, $C_h=dE_h/dT_h$ is given by \cite{us2010}

   \beq
C_h=\frac{2\pi r_h}{g'(r_h)+g''(r_h)r_h}
                                                                        \label{heat}
   \eeq

The cosmological horizon $r_h$ of an observer in the R-region
$0\leq r < r_h$, is the boundary of his manifold, and the second
law of horizons thermodynamics reads $dS_h\geq 0$ (\cite{entropy}
and references therein). It requires $dr_h\geq 0$, so that the
horizon moves outwards. Increasing of a cosmological horizon leads
to decreasing of the mass parameter $M$
\cite{teitelboim,us2010,entropy}.

One-horizon spacetime with $M>M_{cr2}$ and three extrema, evolves
during evaporation towards the double-horizon spacetime with
$M=M_{cr2}$ with zero temperature by virtue of (\ref{temp}). It is
thermodynamically unstable since in a maximum $g'=0$, $g''<0$ and
hence $C_h <0$.

\newpage

Now let us consider a one-horizon spacetime with the metric
function shown in Fig.3 which does not have extrema in the
T-region $r>r_h$ but an inflection point distinguished by two
conditions: $g^{\prime}(r_i)=0; g^{\prime\prime}(r_i)=0$.
\begin{figure}[htp]
\centering
\epsfig{file=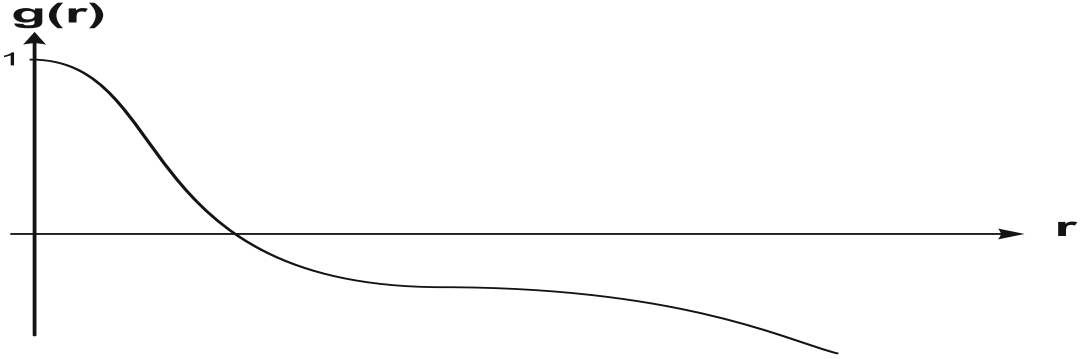,width=14.3cm,height=4.7cm}
\caption{Metric function for one-horizon case with the inflection
point.} \label{fig.4}
\end{figure}

For one-horizon spacetime with the inflection point, specific heat
of the horizon is always positive. Its sign is the same as the
sign of its denominator $g'+g''r_h$. Taking into account
(\ref{trans}) written in the units where $G=1$, we get
$g'+g''r_h=16\pi p_{\perp}r_h-g'$. In a region where
$p_{\perp}\geq 0$, denominator is positive since $g'\leq 0$
everywhere; in a region where $p_{\perp}\leq 0$, eq.(\ref{trans})
gives $g''r_h\leq -2g'$ and thus $g'+g''r_h\leq -g'$ with $-g'\geq
0$, hence $C_h\geq 2\pi r_h/|g'|$.

Evaporation of the cosmological horizon in this case goes towards
the point where decreasing mass achieves a certain critical value
$M=M_{cr}$ at which the metric function vanishes, which
corresponds to the triple horizon $r_h=r_t$ where  the metric
function, its first and second derivatives vanish (Fig.4)
\cite{us2010,entropy}.

\begin{figure}[htp]
\centering \epsfig{file=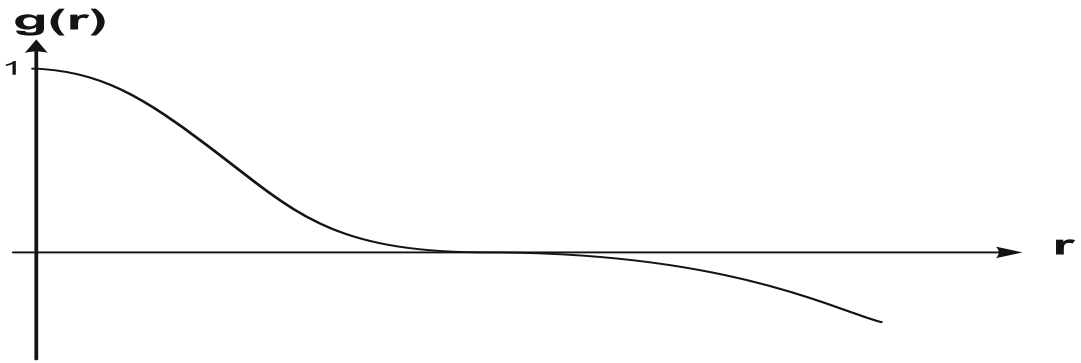,width=14.3cm,height=4.7cm}
\caption{Metric function for the triple horizon spacetime.}
\label{fig.5}
\end{figure}

The specific heat tends to infinity at the triple horizon which
testifies that the triple-horizon spacetime is thermodynamically
stable final product of evaporation of the cosmological horizon in
the course of evolution of a one-horizon spacetime with an
inflection point.

\vskip0.1in

All above analysis was done for an arbitrary density profile with
the only condition imposed on the stress-energy tensor: the weak
energy condition needed to have de Sitter center which in the
cosmological context corresponds to the first inflation. The basic
generic features of the triple horizon $r_h=r_{t}$ are the finite
entropy, zero temperature, zero transversal pressure, zero
curvature, and infinite positive specific heat. Applying the
universal bound (\ref{scramble}), we can qualify  the triple
horizon as extremely long scrambler.

\newpage

 Let us emphasize that the triple-horizon spacetime is not just one of
 possibilities to describe  vacuum dark energy by some one-horizon spacetime.
  - It is distinguished by the quantum dynamics
  of the horizon as the only thermodynamically stable final product of its
evaporation. Moreover, evaporation stops completely ($T_h=0;
~C_h\rightarrow\infty$) at a finite non-zero value of a vacuum
density.  The value of the cosmological constant is tightly fixed
by the final point in the  evaporation of the cosmological
horizon. The triple horizon satisfies three algebraic equations:
$g(r_t)=0; ~ g^{\prime}(r_t)=0;~ g^{\prime\prime}(r_t)=0$, which
define uniquely $M_{cr}$, $r_{t}$, and $q_{cr}$. We have to choose
only the initial vacuum scale $\rho_0$, and then
$g_{cr}^2=\rho_0/\rho_{\lambda}$ will give us the value of
$\rho_{\lambda}$.

\vskip0.1in

Now we can adopt some density profile to check whether the
preferred by horizon dynamics value of the cosmological constant
is close to its observed value.

We choose the density profile \cite{me92} which was used to
produce the above pictures illustrating generic behavior of
spherically symmetric spacetimes with vacuum dark fluid,
   \beq
\rho=\rho_0\exp{(-r^3/2GMr_0^2)}
                                                            \label{profile}
   \eeq

We adopt $\rho_0=\rho_{GUT}$ and $E_{GUT}= 10^{15}$ GeV
 (then $\rho_{GUT}=5\times 10^{77}g~cm^{-3}$ and $r_0=2.4\times 10^{-25}$
 cm),
 and calculate $r_{t}$, $M_{cr}$, and $q_{cr}$, which gives
 $r_{t}=5.4\times 10^{28}$ cm, $M_{cr}=2.33\times 10^{56}$ g, and
$q_{cr}^2=1.37\times 10^{107}$.  We obtain remarkable coincidence
of the calculated parameter $q_{cr}^2$ with that corresponding to
our universe with $\rho_{\lambda~(obs)}\simeq{3.6\times 10^{-30}}
g~cm^{-3}$ \cite{lambda} which gives $q^2=
{\rho_{GUT}}/{\rho_{\lambda~(obs)}}=1.39\times 10^{107}$. (The de
Sitter horizon corresponding to $\rho_{\lambda}$ is $l=9\times
10^{28}$ cm.)

\vskip0.2in

We found that all essential information is stored on the
cosmological horizon and determines its quantum evolution in
accordance with the spirit of the holographic principle.
 We would like to hope that the reduced symmetry of spacetime
 with the triple horizon with an infinite scrambling time
could be also more friendly to a relevant CFT correspondence than
the full symmetry of the de Sitter spacetime and fast-scrambling
character of its horizon.

\subsection*{Acknowledgment}

  This work was supported by the Polish National Science Center through the grant
5828/B/H03/2011/40.

\end{document}